\documentstyle[eqsecnum,multicol,aps,psfig]{revtex} 
\draft
\headheight=\baselineskip
\begin{document}
\bibliographystyle{prsty}
\newcommand{\beq}{\begin{eqn}}
\newcommand{\eeq}{\end{eqn}}
\title{A reaction-diffusion model for the hydration
of calciumsulphate (gypsum) and microstructure percolation.}
\author{F.~Tzschichholz$^{1}$ and H.~J.~Herrmann$^1$}
\address{
$^1$ Institut f\"ur Computeranwendungen I,
Universit\"at Stuttgart,\\
Pfaffenwaldring 27,
D--70569 Stuttgart, Germany}
\date{\today}
\maketitle
\begin{abstract} 
We have numerically investigated a reaction-diffusion model for the 
hydration of calciumsulphate (gypsum). The simulations were conducted for 
two and three dimensional systems. While the dissolution of 
anhydrous gypsum is considered irreversible at a finite rate
the precipitation/dissolution
reaction for the calciumdihydrate is considered reversible. The latter
reaction is assumed to be controlled by the dihydrate's equilibrium 
solubility {\em and} the abillity of the system to react on 
supersaturation only at a certain velocity described by the reaction 
rate constant of precipitation.
For $d=2$ we find at early times an accelerated hydration period followed 
by a maximum and a decreasing hydration rate. For large times 
the ionic product of involved species assumes closely the value 
of the di-hydrate equilibrium solubility. Calculated model 
micro-structures exhibit typical features such as inner and outer 
hydrate products, induction and dormant period as well as bridging.
Furthermore we find that the overall chemical reactivity as a function
of initial anhydrous (volume) concentration $p$
exhibits a maximum close to the percolation point of the underlying 
lattice.  
Employing a rescaling procedure we find {\em two} percolation thresholds 
in $d=2$,
$p_c^{min}=0.44\pm 0.015$ and $p_c^{max}=0.77\pm 0.02$,  for the 
initial anhydrous gypsum concentration {\em between} which 
percolating dihydrate structures can be attained. For $d=3$ we find 
$p_c^{min}=0.10\pm 0.02$ and $p_c^{max}=0.95 \pm 0.02$.
\end{abstract} 

\begin{multicols}{2}

\section{Introduction}\label{sec:Introduction}
Gypsum is perhaps one of the oldest crafting and building materials 
human kind has cultivated beside wood, iron, and stone. 
There exists a long tradition 
of using this material in arts, medicine, paleontology,
archeology, as well
as a light-weighted building material.
Its widespread use is likely due to its natural abundance, its almost total
flexibility in applications, its chemical inertness and certainly its 
low costs. For an overview of the various applications we refer 
the interested reader to the survey of Wirsching\cite{ullstein}\\
Gypsum belongs to the group of calcium-sulphates with its various 
hydrates. A material class exhibiting some common phenomenological 
properties are the calcium-silicates, mostly termed as 
cementious materials. Cementious materials are, however, of much 
higher intrinsic strength than gypsum due to their chemical shrinkage 
and are therefore in technical applications mostly prefered to gypsum. 
While most cementious materials do precipitate as a 
gelantineous phase, calcium-sulphates appear always in crystalline 
modifications (needles).\\
Both calcium-sulphates as well as calcium-silicates are 
already well characterized on a molecular level\cite{Taylor90,Kuzel87}.
However, it appears to us that gypsum 
is less understood compared to calcium-silicates on the macroscopic scale. 
For more recent experimental investigations we refer the interested
reader to Ref.~\ref{Lewry94c}. 
For cementious materials microstructure computer models have been 
recently devellopped\cite{Bentz91a,Bentz94a,Made94,thz96}. Though the 
cementious systems are much more complex from a chemical point of view
their huge economical importance as the building material of choice 
has focussed the attention on the modeller's side.
In the current work we would like to present some first results 
for the microstructure of hydrated gypsum
based on diffusion-reaction calculations.
In the practice, the most important feature of the setting of gypsum is the 
formation of rigid structure of the hydrates. The necessary condition 
for this to occur is a connected structure, in other words, that 
the hydrate aggregate percolates. In this paper we will therefore investigate
under which conditions (concentration of a hydrate) the end 
product of the reaction forms a percolating cluster.

\section{Physicochemical Aspects}\label{sec:physicochemistry}
The hydration of anhydrous calciumsulphate,$CaSO_4$ - in the following $CS$, 
is based on dissolution/precipitation reactions forming various 
calciumsulfohydrates\cite{ullstein}.
The hydrates, $CaSO_4\cdot n\, H_2O$, are distinguished by the molar amount 
$n$ of bound crystal-water. In connection with the amount of physically 
bound water one observes different physico-chemical properties as for example 
crystal-symmetries, densities, or solubilities.
The two most important hydrates are the semihydrate, $n=1/2$, and the 
dihydrate, $n=2$. There are two 'modifications' of the semihydrate called the
$\alpha-$ and $\beta-$semihydrate. On the side of the anhydrous 
calciumsulphates three forms are distinguished: soluable CS (A III),
dead CS (A II), and high temperature CS (A I). Typical solubilities at 
room temperature and normal pressure are     $2\,gramm/liter$ for the 
dihydrate, $2.5\,gramm/liter$
for A\,II, $5.8\,gramm/liter$ for $\alpha$-semihydrate, and 
$7.6\,gramm/liter$ for $\beta$-semihydrate. 
The solubilities for $\alpha-$,$\beta-$semihydrate and A\,II decrease 
monotonically with increasing temperature while the dihydrate exhibits 
a flat maximum at temperatures $T=40\; Celsius$. 
Most minerals containing $CS$ are grouped into three categories of materials a)
gypsum which consists almost only of dihydrate, b) anhydrous stone 
consisting almost only of A\,II, and c) technological mixing forms 
consisting mostly of A\,II and  A\,III and semihydrates.\\

Stochiometrically one has the following net reactions:\\
{\em hydration of anhydride to dihydrate}
\begin{equation}
CaSO_4 + 2H_2O \to CaSO_4\cdot 2H_2O,
\end{equation}  
{\em hydration of semihydrate to dihydrate}
\begin{equation}
2CaSO_4\cdot \frac{1}{2}H_2O +3H_2O \to 2CaSO_4\cdot 2H_2O.
\end{equation}
The implementation of the foregoing reactions in a reaction-diffusion model
requires explicit knowledge of the dissolution/precipitation reactions, i.e.,
their kinetic rate equations.
Observing that these rate equations are not explicitely known to us we make 
the following assumptions.
\begin{enumerate}
\item {\em Solubilities.} 
The present calciumsulphate system is strongly electrochemical.
In the following we will assume nevertheless that all electrostatic 
interactions 
between solvated ions can be disregarded. Solubilities are therefore given
by ionic products of concentrations. In particular,  
$S_{CaSO_4}^{(\infty)} \approx 10^3 Gramm/liter  \gg$
$S_{CaSO_4\cdot 2\;H_2O} \approx 2 Gramm/liter$,
$S_{CaSO_4\cdot \frac{1}{2}\;H_2O} \approx 6,5 Gramm/liter$.
All values are equilibrium values at room temperature and normal pressure.
The value for $S_{CaSO_4}^{(\infty)}$ is supposed to correspond to an 
infinitely high solubility.
\item {\em Diffusion constants.}
Correspondingly we employ diffusion constants as observed for 
'infinite dilution' neglecting all possible electrochemical influences.
This is a simplification for liquid electrolytes.
In particular, $D^{Ca^{2+} }\approx 7.9\cdot 
10^{-10}\;m^2/s$, $D^{SO_4^{2-} }\approx 10.7 \cdot 10^{-10}\;m^2/s$.

\item {\em Solvation.}
Considering the process of solvation it becomes clear that the above 
hydration reaction needs to be split up at least into a dissolution and a 
precipitation reaction. The formed ions are solvated ions {\em binding} 
a certain amount of water. Just let us write down the dissolution reaction,
\begin{eqnarray}
\label{dissolution_eq}
(CaSO_4)_{(s)} +(\nu_{Ca}+\nu_{SO_4})H_2O_{(\ell)} \to \\ 
(Ca^{2+},\nu_{Ca}H_2O)_{(aq)} +  (SO_4^{2-},\nu_{SO_4}H_2O)_{(aq)}, 
\end{eqnarray}
und the precipitation reaction,
\begin{eqnarray}
\label{precipitation_eq}
(Ca^{2+},\nu_{Ca}H_2O)_{(aq)} +  (SO_4^{2-},\nu_{SO}H_2O)_{(aq)} \to \\
(CaSO_4\cdot 2H_2O)_{(s)} +(\nu_{Ca}+\nu_{SO_4}-2)H_2O_{(\ell)}.
\end{eqnarray}
The numbers $\nu_{Ca}$ and $\nu_{SO_4}$ do characterize the formed 
hydrate shells of the calcium- and sulphate-Ions. 
One has $\nu_{Ca}=6$. We are currently not aware 
of the value for $\nu_{SO_4}$. 
For the simulations we tentatively assume $\nu_{SO_4}=6$.

\item {\em Specific molecular volumes.}
The specific volumina of the involved solid and liquid phases under normal 
conditions were obtained
from the literature\cite{ullstein}, in particular:
$v_{CaSO_4}= 52.7 \cdot 10^{-3} liter/mol$, 
$v_{CaSO_4\cdot 2H_2O}= 74.1 \cdot 10^{-3} liter/mol$, 
$v^{(\alpha)}_{CaSO_4\cdot 0.5H_2O}= 55,7 \cdot 10^{-3} liter/mol$, 
$v^{(\beta)}_{CaSO_4\cdot 0.5H_2O}= 52.7 \cdot 10^{-3} liter/mol$ 
and $v_{H_2O}=18 \cdot 10^{-3} liter/mol $.

\item {\em Reaction constants.}
The reaction constants were chosen such that firstly the back reaction of the 
anhydrous dissolution becomes negligibly small (irreversible reaction) 
and secondly the precipitation/dissolution of dihydrate is adjusted 
to the experimentally observed value for dihydrate solubility.
For {\em calciumsilicates} typical scales for the involved surface reaction 
rates are $10^{-6}\frac{m^4}{mol\cdot s}$.
We have tried to orientate ourselves on these magnitudes. 
\end{enumerate} 

\section{The Model}\label{sec:Model}
In the following we consider only the hydration of anhydrous gypsum 
towards dihydrate. As precise reaction mechanisms and rates are not 
known to us it appears reasonable to model just one dissolution and one 
precipitation reaction, i.e., Eqs.~(\ref{dissolution_eq}) and 
(\ref{precipitation_eq}).
The general model setup has been described elsewhere\cite{thz96}. 
For the presented calculations we employed a time integration step 
$\Delta t=10^{-1}\;s$ and a spatial resolution $\Delta x=10^{-4}\;m$.
The calculations were performed in $d=2$ for a system of size 
$100\times 100$ and in $d=3$ for sizes of $50\times 50 \times 50$. 
The dissolution rate constant was prescribed and fixed 
at $k_{diss}=10^{-3}\;\frac{mol}{m^2\,s}$ (absolute scale) for 
a somewhat lower precipitation-rate-constant 
$k_{prec}=10^{-2}\cdot k_{diss}$. 
The by a factor $100$ lower precipitation rate constant was choosen in 
order to mimick the experimental observation, of a much faster 
dissolution than precipitation.

\section{Results}\label{sec:Results}
\subsection{Results for $d=2$}\label{subsec:Results_d=2}
\begin{figure}[tb]
        \centerline{    
        \psfig{file=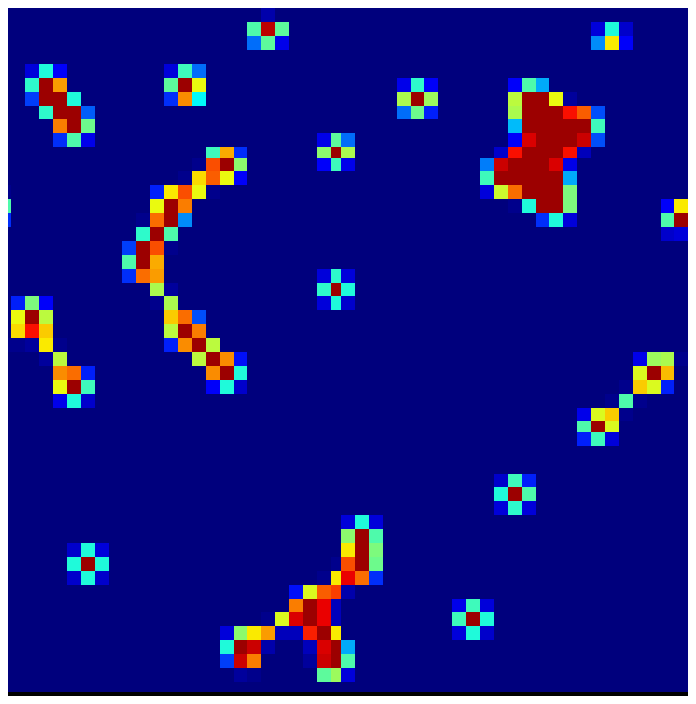,width=0.25\textwidth}
        \psfig{file=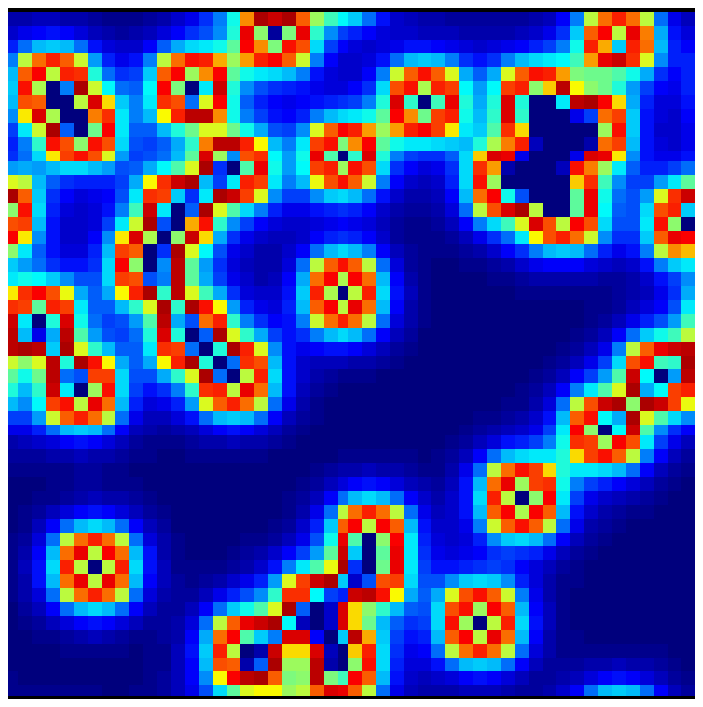,width=0.25\textwidth}
        \psfig{file=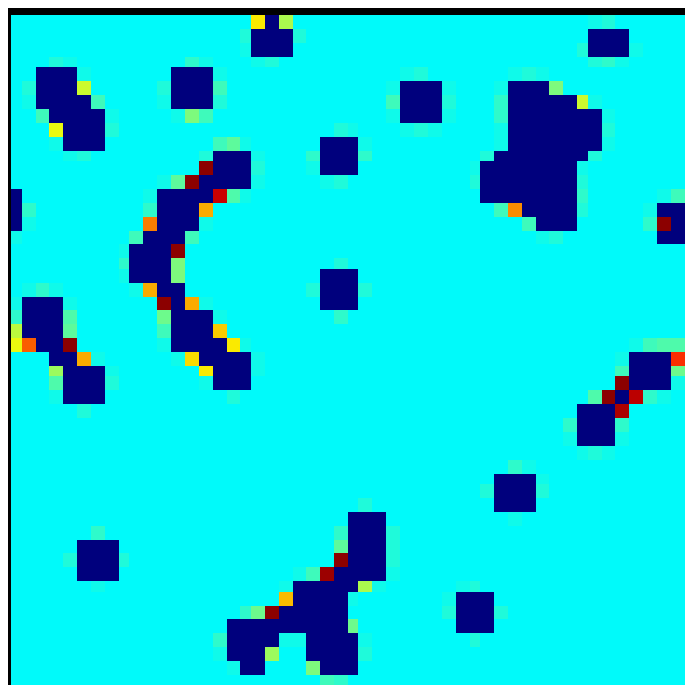,width=0.25\textwidth}
}
        \caption{
        Volume concentrations for anhydrous gypsum, dihydrate, 
        and ions (left to right). Red represents high - blue
        low concentrations. $k_{diss}=10^{-3}\;\frac{mol}{m^2\,s}$,
        $k_{prec}=0.1\cdot k_{diss}$, $\Delta t=10^{-1}\;s$, 
        $\Delta x=10^{-4}\;m$, $L=50$, Time $t=8\cdot 10 ^4s$, 
        average ion concentration $12\cdot 10^{-3}\frac{mol}{liter}$.
}
        \label{struc_2d_kprec<kdiss}
\end{figure}

Fig.~\ref{struc_2d_kprec<kdiss} represents the volume concentrations 
for anhydrous gypsum, dihydrate, and ions (left to right). 
Red means high - blue low concentrations.
\end{multicols}
\begin{multicols}{2}

\end{multicols}
\begin{figure}[tb]
        \centerline{
        \psfig{file=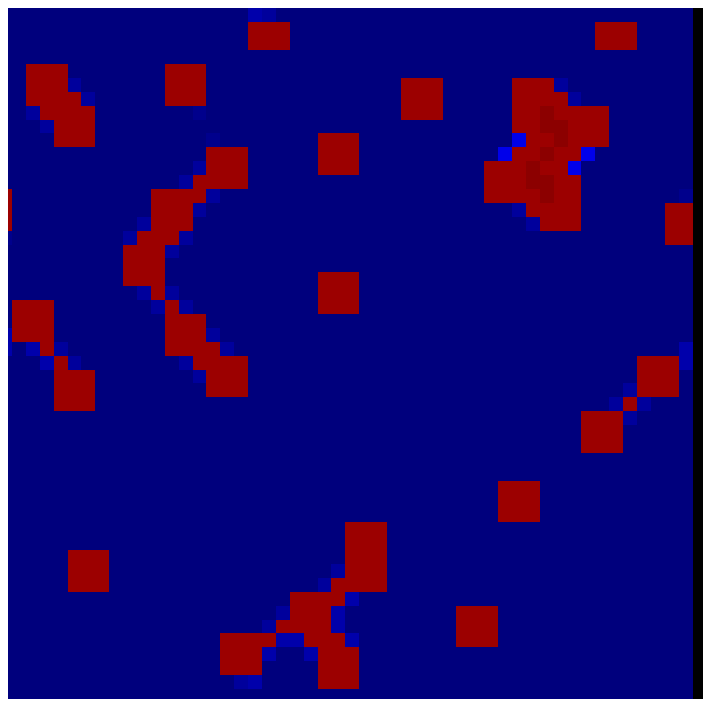,width=0.25\textwidth}
        \psfig{file=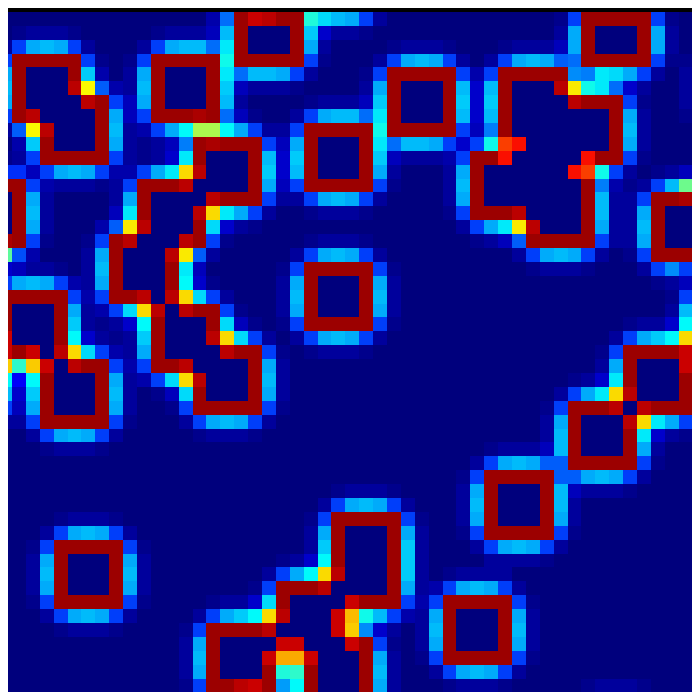,width=0.25\textwidth}
        \psfig{file=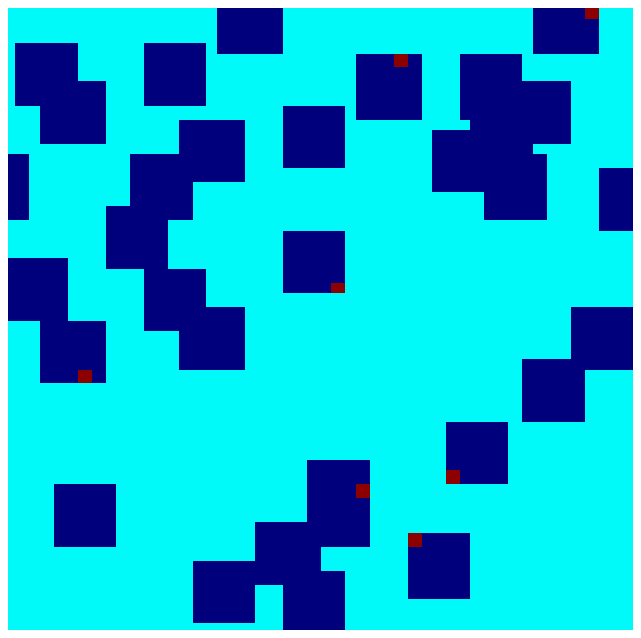,width=0.25\textwidth}
}
        \caption{
        Volume concentrations for anhydrous gypsum, dihydrate, 
        and ions (left to right). Red represents high - blue
        low concentrations. $k_{diss}=10^{-3}\;\frac{mol}{m^2\,s}$,
        $k_{prec}=k_{diss}$, $\Delta t=10^{-1}\;s$, 
        $\Delta x=10^{-4}\;m$, $L=50$, Time $t=8\cdot 10 ^4s$, 
        average ion concentration $11.6\cdot 10^{-3}\frac{mol}{liter}$.
}
        \label{struc_2d_kprec=kdiss}
\end{figure}
\begin{multicols}{2}
\end{multicols}
\begin{figure}[tb]
        \centerline{
        \psfig{file=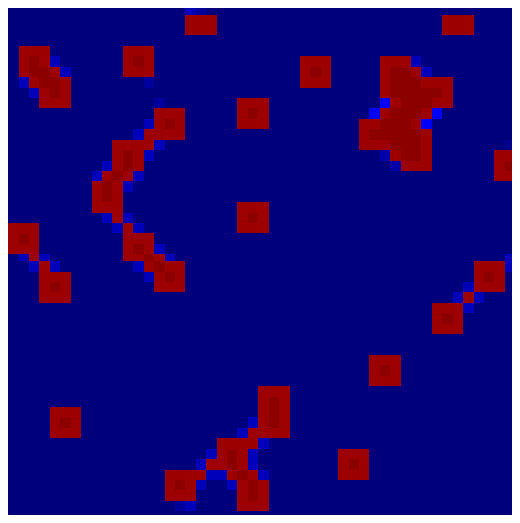,width=0.25\textwidth}
        \psfig{file=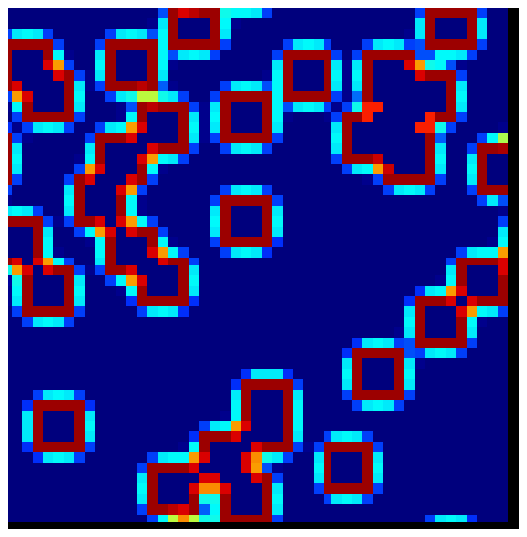,width=0.25\textwidth}
        \psfig{file=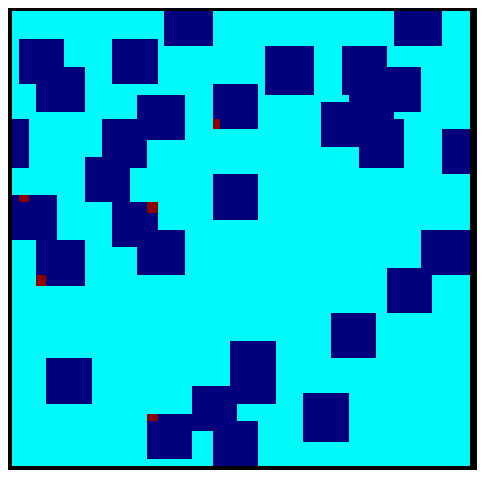,width=0.25\textwidth}
}
        \caption{
        Volume concentrations for anhydrous gypsum, dihydrate, 
        and ions (left to right). Red represents high - blue
        low concentrations. $k_{diss}=10^{-3}\;\frac{mol}{m^2\,s}$,
        $k_{prec}=10\cdot k_{diss}$, $\Delta t=10^{-1}\;s$, 
        $\Delta x=10^{-4}\;m$, $L=50$, Time $t=8\cdot 10 ^4s$, 
        average ion concentration $11.6\cdot 10^{-3}\frac{mol}{liter}$.
}
        \label{struc_2d_kprec>kdiss}
\end{figure}
\begin{multicols}{2}

The cases $k_{prec}=0.1\cdot k_{diss}$ and $k_{prec}=k_{diss}$
exhibit similarities where the early hydration undergoes an 'induction' 
period. In this induction period the ion concentrations rise steeply due 
to a relatively high anhydrous gypsum dissolution.   
After passing a maximum ion concentration the main reaction period 
begins to emerge in which 'outer' dihydrate starts to precipitate
blobking further reactions of the anhydrid.
Correspondingly ion concentrations drop to an almost stationary value, 
considerably slowing down the whole further hydration in the vicinity 
of the anhydrous gypsum grains. 
Such behavior is also known from calciumsilicates during the 
formation of inner hydrates (experimentally and numerically).
The for $k_{prec}$ over two decades almost constant ion concentration 
corresponds
essentially to the equilibrium solubility for dihydrates of about 
$11.6\cdot 10^{-3}\frac{mol}{liter}$. 
As the solvent is slightly supersatured with ions the solvated anhydrous 
gypsum is precipitated relatively quickly again.
The velocity with which the precipitation reaction reacts with respect 
to the supersaturation is controlled by the rate constant  $k_{prec}$.

In order to study the influence of the initial volume of anhydrous gypsum
$p$ on the hydration process a set of calculations was conducted.
The calculations were performed for $k_{diss}=10^{-3} \frac{mol}{m^2 s}$ 
and $k_{prec}=10^{-2} k_{diss}$. The basic investigated question was 
for which $p$ critical percolation of the {\em hydrated phase} in the 
long run occurs.\\

\begin{figure}[tb]
        \centerline{
        \psfig{file=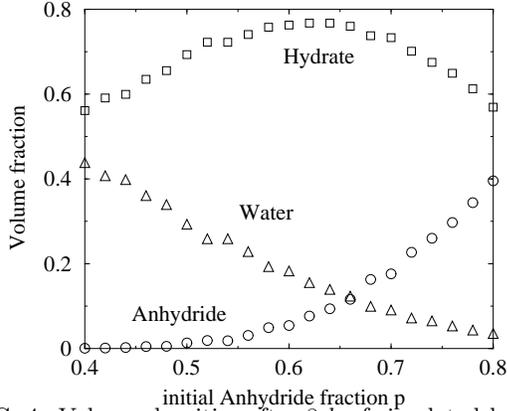,width=0.40\textwidth}
}
        \caption{
        Volume densities after $8\;h$ of simulated hydration.
        $k_{diss}=10^{-3} \frac{mol}{m^2 s}$, $k_{prec}=10^{-2} k_{diss}$,
        $L=100$, anhydrate ($\circ$), water ($\triangle$), 
        dihydrate ($\Box$). 
        All other parameters as 
        in Fig.~\ref{struc_2d_kprec<kdiss}.     
}
        \label{volumedensity_2d}
\end{figure}
As a first investigation the total volume densities after eight 
hours of hydration time were measured in a range between 
$p=0.4$ and $p=0.8$, see Figure \ref{volumedensity_2d}. 
The system size was $100 \times 100$.
The numerical effort per data point was about 40 minutes 
cpu time on a 2GHz PC.
     
For initial anhydrous gypsum densities less than $p=0.54$ the final 
remaining gypsum is less than $0.02$, i.e., the dissolution reaction is 
complete. Above $p=0.6$ one observes a strongly increasing fraction
of non dissolved gypsum, i.e., the reactions become increasingly incomplete.
As water is not treated as limiting reactant {\em per se} in the calculations
there must exist a non-stochiometric reason for the incomplete reaction.
The amount of precipitated dihydrate $(\Box)$ confirms this.
The corresponding curve undergoes a maximum in reactivity for $p$ values 
between $0.6$ and $0.65$ although sufficient amounts of water are 
{\em system wide} available.
This effect is most likely due to a screening of the initial 
anhydrous gypsum.\\
Above $p_c= 0.592$ increasingly parts of the chemically reactive interface 
between water and gypsum are blocked by the gypsum itself.
For $p > p_c$ one has to expect a decreasing specific surface area for the 
gypsum - which corresponds directly to a decreasing chemical reactivity.
Let $p$ denote the total volume fraction of 
initial anhydrous gypsum.  
As the hydrate density is a more or less continous quantity 
$\rho (x,p)\in [0,1]$ one has to term the statement of percolation 
somewhat different than 'usual'. 
Looking at the densities $\rho$, we define a function $R(x,p,r)$
with the property,

\begin{equation}
R(x,p, r)=\Theta (\rho (x,p)-r) 
\end{equation}

(with $\Theta$ being the Heaviside step function). 
The function $R(x,p,r)$ maps just all those volume elements to unity (red) 
for which the corresponding hydrate density exceeds a certain amount $r$, 
otherwise it becomes zero (blue).
The following images show the hydrate pixels 
$R(x, r)$ at $r=0.7$ for $p=0.4$, $p=0.5$, $p=0.7$, and $p=0.8$ (from left 
to right).\\

\end{multicols}
\begin{figure}[tb]
        \centerline{
        \psfig{file=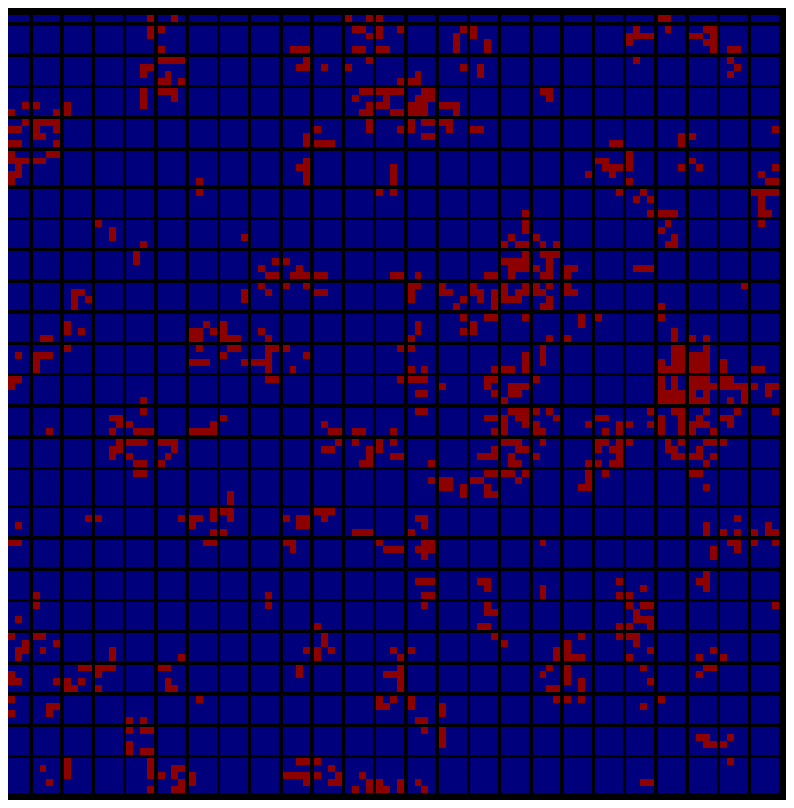,width=0.25\textwidth}
        \psfig{file=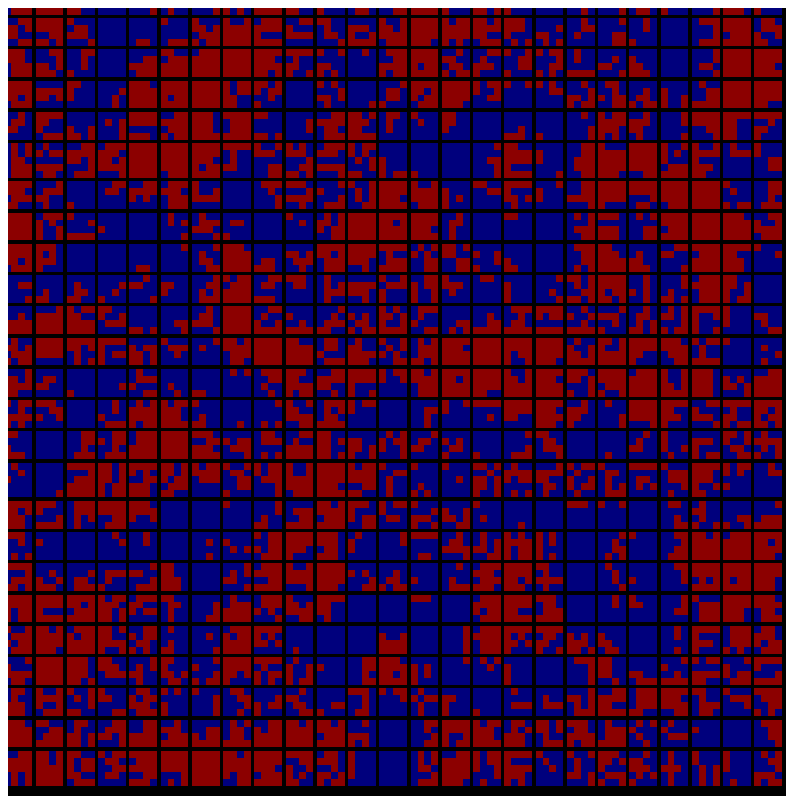,width=0.25\textwidth}
        \psfig{file=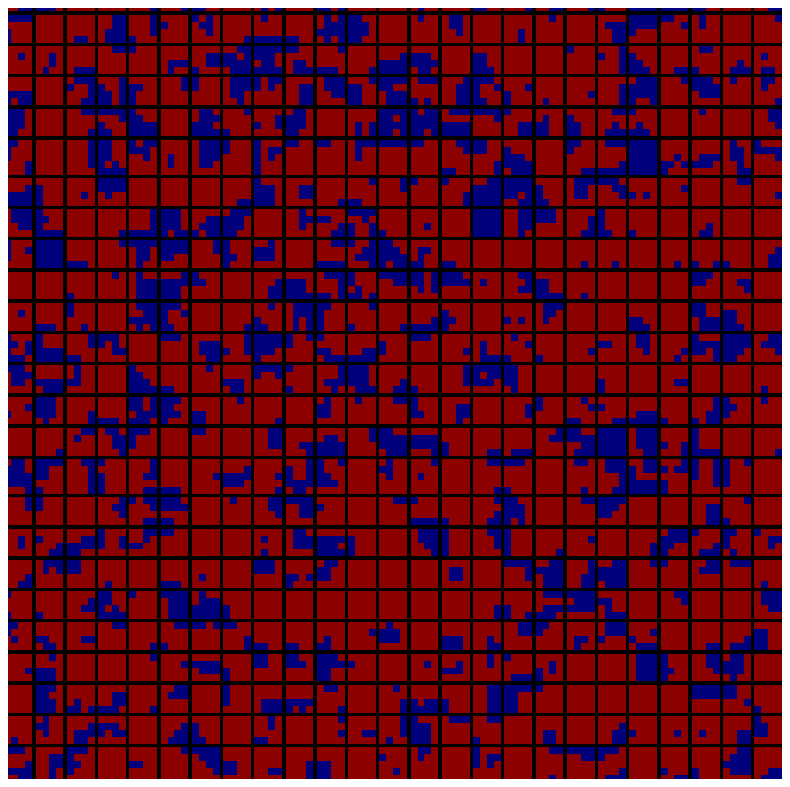,width=0.25\textwidth}
        \psfig{file=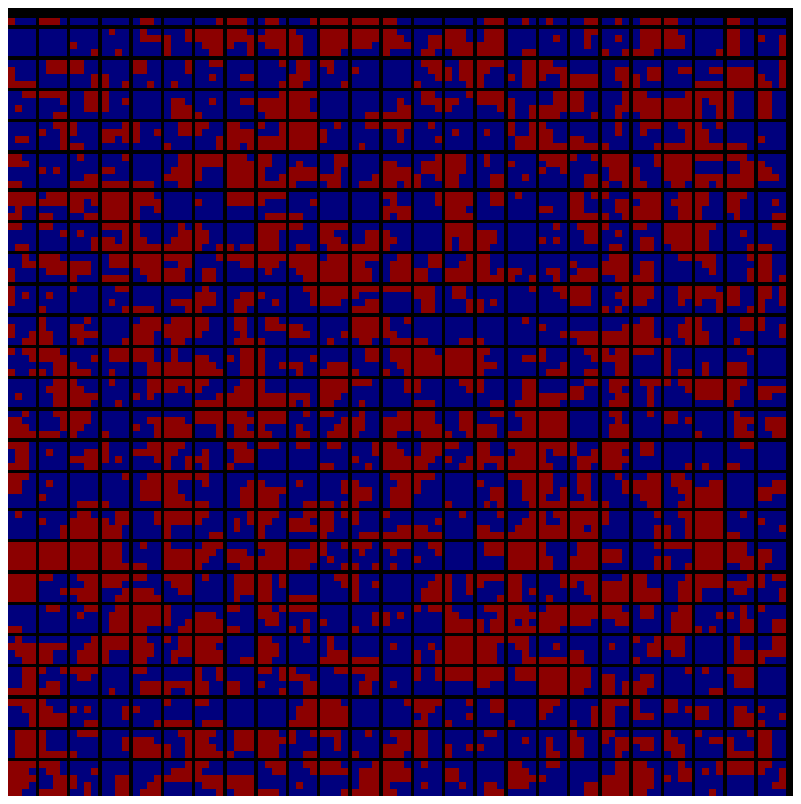,width=0.25\textwidth}
}
        \caption{
        Hydrate pixels at $r=0.7$ for initial anhydrate
        volume concentrations $p=0.4$, $p=0.5$, 
        $p=0.7$, and $p=0.8$ (from left to right). A red pixel corresponds 
        to an asymptotic dihydrate volume concentration larger than $r$
        whereas a blue pixel represents a concentration lower than $r$.
        Parameters as in Fig.~\ref{volumedensity_2d}.
}
        \label{struc_2d_filtered}
\end{figure}
\begin{multicols}{2}

From these images it is apparent that for very high gypsum densities, i.e., 
$p=0.8$, the hydrate does not percolate that easily as compared for 
example to $p=0.6$. This indicates that at least two percolation transitions 
for the hydrated phase do exist.\\

We employ volume concentrations in this work.
Obviously the solid phase concentrations define a 
measure for the distinct properties. How to define criteria 
in terms of phase concentration(s) that allow an analogy to percolation
phenomena? 
In conventional (site-)percolation each volume element is either 
completely occupied by species $A$ or $B$ at random in an {\em uncorrelated} 
way with fixed occupation probabilities $p_A$ and $1-p_A$ respectively.
The constant occupation probability $p_A$ equals the overall density of
species $A$, and the formed cluster at the percolation transition exhibits
scale invariance (selfsimilarity)\cite{burning}. 
Vice versa one could employ this scale invariance as a tool in order to 
estimate the critical percolation probability $p_A$.
We will follow this approach in the present paper. 
Because the hydrate concentrations directly correspond to non-constant 
occupation probabilities one could expect {\em correlated}
percolation. 

The microstructural and topological properties 
of the transformed phase strongly influence overall (volume) properties
such as elastic properties. If -as in our case- the unreacted phase
owns a small or even vanishing elastic modulus 
(anhydrous gypsum is usually a powder) then the percolating phase controls
entirely overall properties. 
Thereforethe percolation transition
of the spatial dihydrate distribution is of central importance to a 
corresponding 'transition' of mechanical properties for the considered 
system. Naturally the question arises on how to estimate and characterize 
a possible percolation transition as for our case. 
We introduce the {\em typical}  
volume density $r$ above which all hydrated volume elements are regarded as 
occupied sites in the sense of percolation. Let us denote $k(p,r)$
as the true overall density of such sites. In general $k(p,r)$ and $r$ will
have different values because the overall density $k$ depends on the anhydrous 
volume fraction $p$ while $r$ is a variable. 
However, if we guarantee by some algorithm
that the occupied sites do {\em always} percolate for arbitrary values 
of $r$ (scales) by adjusting the initial $p$, then the overall density $k$ 
must equal the typical density $r$. In other words $k=r$ 
which is a fixed point of the map $k(p,r)$.

In order to investigate this question quantitatively more extensive 
calculations have been performed.
One can calculate from the pixel-distribution $R(x,p,r)$ two main quantities 
of interest a) its overall density,
\begin{equation}
k(p,r)=\frac{1}{V}\int d^3x\; R(x,p,r)
\end{equation}

and b) the information whether {\em the pixel-distribution} $R$ is percolating 
itself (via a burning algorithm\cite{burning}).
The set of rules {\em without} the fixed point condition is the following:
\begin{enumerate}
\item Choose an initial $r=0.593$, 
\item Choose a sufficient low initial gypsum 
density, i.e., $p=0.3$ in order to find later on $p_{min}$.
\item Do the simulation until stationary concentration fields are reached 
( here $t=3\cdot 10^4\;s$).
\item Calculate $R(x,p,r)$ and $k(p,r)$. Determine whether $R(x,p,r)$ is 
percolating or not, and choose accordingly a decremented/incremented $p$.
A hyperbolic decreasing interval increment/decrement was chosen.
\item Continue with 3. until the increment interval becomes lower than 
$\delta p = 10^{-3}$.
\item Choose a different initial value for $r$, and continue with 2.
\end{enumerate} 
The algorithm works similar for $p_{max}$.

\begin{figure}[tb]
        \centerline{    
        \psfig{file=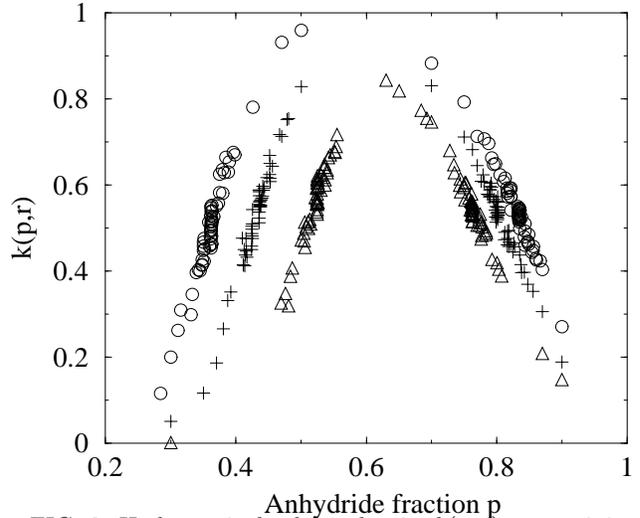,width=0.5\textwidth}
}
        \caption{
        Hydrate pixel volume density,$k(p,r)$, versus 
        initial anhydrate concentrations, $p$, for three different 
        cutoffs $r=0.5 (\circ)$, $r=0.593 (+)$ and $r=0.7 (\Delta)$.
        Parameters as in Fig.~\ref{volumedensity_2d}.   
}
        \label{p_k_2d}
\end{figure}

Fig.~\ref{p_k_2d} shows some typical plots of hydrate densities
$k(p,r)$
and corresponding initial anhydrate concentrations for three different $r$. 


The data points link the dependence of the density $k$ and 
initial anhydrous volume density $p$ for various cutoffs $r$ under the 
condition that $R$ is percolating.

\begin{figure}[tb]
        \centerline{    
        \psfig{file=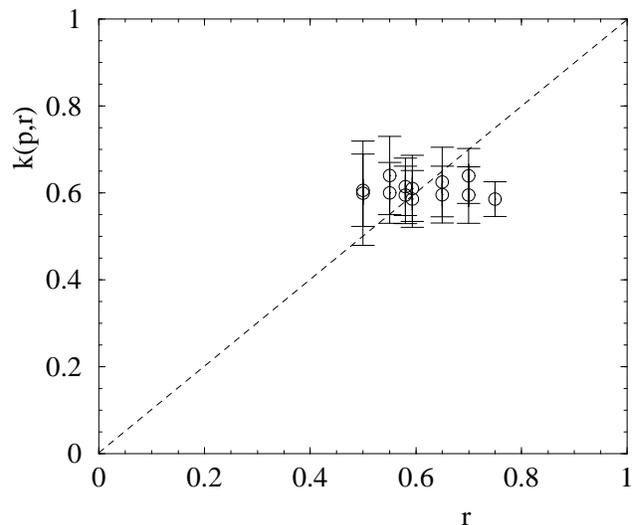,width=0.50\textwidth}
}
        \caption{
        Averaged hydrate pixel volume density, $k(p_c,r)$, versus 
        cutoffs $r$ for the various critical anhydride densities 
        as observed from Fig.~\ref{p_k_2d}.
        Points that intersect the line $k(p_c,r)=r$ define a critical value
        $k_c(p_c,k_c)$.
        Parameters as in Fig.~\ref{volumedensity_2d}. 
}
        \label{p_k_with_error_bars_2d}
\end{figure}

It is more meaningful to average {\em all} points for the same $r$ 
belonging to a percolating cluster, and calculate their mean and variance 
values, which results in one single point with error bars per value of $r$. 
This is shown in Fig.\ref{p_k_with_error_bars_2d}.

We now impose the condition $k(r,p_c)=r$ on the data points shown in 
Fig.~\ref{p_k_with_error_bars_2d}.
We find a common {\em critical} value $k_c=0.62     \;(+0.03/-0.04)$.
From the lowest and highest value for $k_c$
one obtains the desired values for the minimum and maximum 
percolation threshold 
$p_c^{min}=0.44  \;(+0.015/-0.015)$ and $p_c^{max}=0.77    \;(+0.02/-0.02)$.
This means in particular that within the error bar of $k_c$
there are no different values for $k_c$ for $p_c^{min}$ und $p_c^{max}$.

\subsection{Results for $d=3$}\label{subsec:Results_d=3}

In the foregoing paragraph we investigated a few percolation properties 
in $d=2$ depending on the initial anhydrous gypsum concentration. 
We found that 
for the case of a diluted ($p < p_c^{min}$) and dense 
initial anhydride packing
($p > p_c^{max}$) no percolation of the hydrated phase will occur, i.e.
the hydrated gypsum becomes useless for most technical applications.
We investigated this problem also in $d=3$. To this end the reaction-diffusion 
model and the burning algorithm were modified. 
\begin{figure}[htb]
        \centerline{    
        \psfig{file=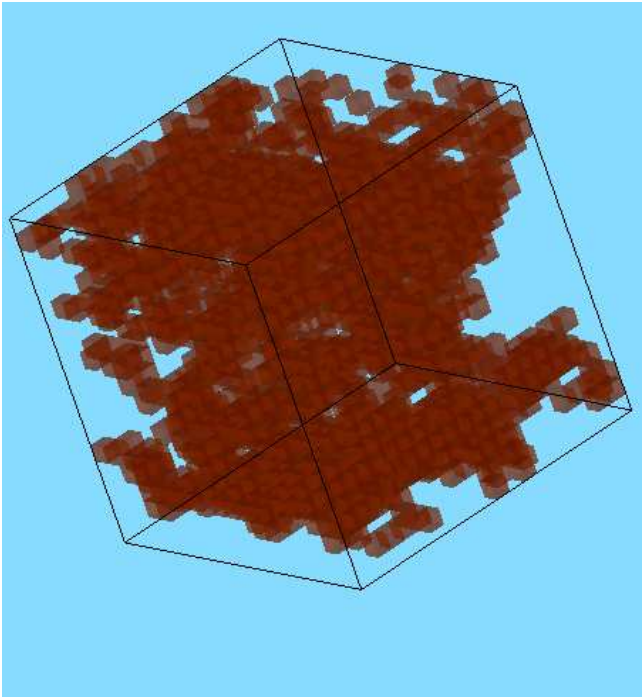,width=0.5\textwidth}
}
        \caption{
        A typical dihydrate backbone in $d=3$, systemsize $20^3$.       
}
        \label{backbone_3d}
\end{figure}
In Fig.~\ref{backbone_3d} we show a small hydrate backbone, i.e. sites not 
belonging to a spanning cluster have been removed. One clearly sees 
the holes in the microstructure constituting the pore space. 
In order to determine the further percolation properties
we employed the same 
numerical procedure and parameters as explained in Sec.\ref{subsec:Results_d=2}
with the exception of the system size being now $50^3$.

\begin{figure}[tb]
        \centerline{    
        \psfig{file=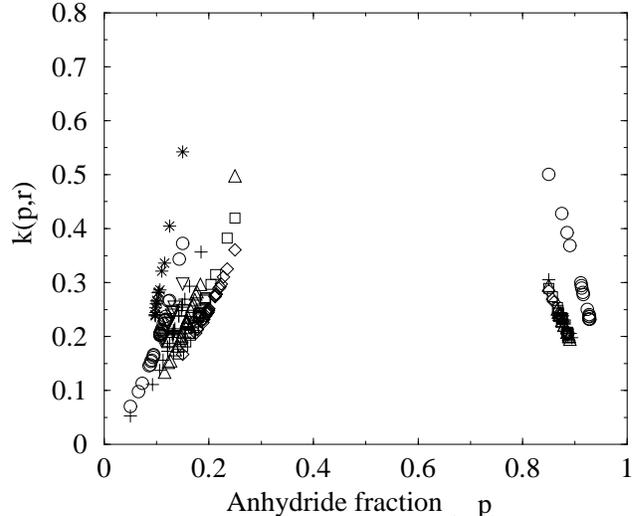,width=0.5\textwidth}
}
        \caption{
        Hydrate pixel volume density,$k(p,r)$, versus 
        initial anhydride concentrations, $p$, for seven different 
        cutoffs $r=0.25(\star)$,
        $r=0.27(\circ)$, $r=0.29(\nabla)$, $r=0.312(+)$, 
        $r=0.33(\Delta)$,
        $r=0.35 (\Box)$ and  $r=0.37(\Diamond) $. 
        Other parameters:
        $k_{diss}=10^{-3} \frac{mol}{m^2 s}$, $k_{prec}=10^{-2} k_{diss}$,
        $L=50$ in $d=3$. 
}
        \label{p_k_3d}
\end{figure}

The numerical effort was about 7 days CPU time on a 2GHz PC per value of $r$. 
In Fig.\ref{p_k_3d} 
we show the $k(r,p)$ characteristics for seven different 
values of $r$. 

\begin{figure}[tb]
        \centerline{    
        \psfig{file=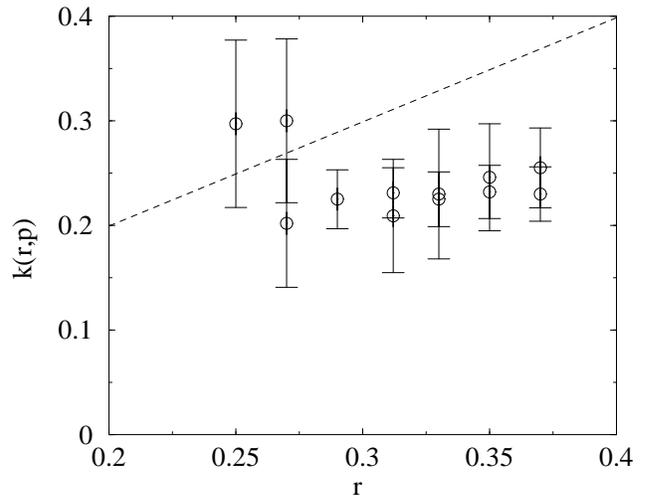,width=0.5\textwidth}}
        \caption{
        Typical volume density, $k(p_c,r)$, plotted versus $r$. Points 
        that intersect the dashed line $k(p_c,r)=r$ define the critical 
        $k_c=0.25\pm 0.02$ in $d=3$.
        Parameters as in Fig.~\ref{p_k_3d}. 
}       
\label{k_r_3d}
\end{figure}

The scatter of data for each $r$ value appears much smaller 
than in Fig.\ref{p_k_2d}. Of course are the limits for $k(r,p\to p_c)$
smaller now, see~Fig.\ref{p_k_2d}, because the thresholds for uncorrelated 
percolation in $d=3$ are much smaller than in $d=2$. In $d=3$ there exist 
much more connected pathways for ion diffusion and solvation.
Therefore it is reasonable to expect that a possible upper percolation 
threshold $p_c^{max}$ is shifted to higher anhydrous concentrations 
as compared to Fig.~\ref{p_k_2d}.
Again in Fig.~\ref{k_r_3d} we show mean and variance 
values for $k$ versus $r$.

Imposing the condition $k(r,p_c)=r$ we obtain
within the error bars $k_c=0.25\pm 0.02$ implying 
$p_c^{min}=0.10\pm 0.02$ 
and $p_c^{max}=0.95\pm 0.02$. As $p_c^{max}$ is very high compared to 
$d=2$ this could play a certain role in the rigidity limit at very 
dense packings (poly-disperse packings). 

\section{Conclusion}
We studied percolation properties of a reaction-diffusion model 
for the hydration of calciumsulphate. From a kinetic point of view 
the basic effects of the dissolving and precipitating microstructures are 
a) that they control the ion diffusion transport to a large extent 
and b) that they form a spanning backbone from one systems end to the other, 
thus allowing the tyransport of momentum/stress. 

We expect that for $p<p_c^{min}$ and $p> p_c^{max}$ microstructures are
unable to carry stress at all. 
It should be noted that for $d=2$ and $d=3$ $p_c^{max}$ is not reachable 
employing mono-disperse packings. 
  
Our results also show that there exists an optimum initial anhydrous
concentration for which the total amount of dihydrate reaches a maximum,
i.e. shows maximum overall chemical reactivity, and possibly the 
best mechanical properties.
  
The relatively high computing time for $d=3$ indicates that one possibly 
should consider in the future
a different evaluation method than exact enumeration in order to solve the 
reaction-diffusion equations.

%
%

\end{multicols} 
\vfill\eject 


\begin{thebibliography}{10}

\bibitem{ullstein}\label{ullstein}
F. Wirsching, {\em Calcium Sulphate} in 
Ullmann's Encyclopedia of Industrial Chemistry, 6th Edition, (1998).


\bibitem{Taylor90}\label{Taylor90}
{\em Cement Chemistry},
H.~F.~W.~Taylor,
Academic Press, 1990.

\bibitem{Kuzel87}\label{Kuzel87}
Kuzel and Hauner, 
Zement Kalk Gips {\bf 40}, 628-32 (1987).

\bibitem{Lewry94c}\label{Lewry94c}
A.~J.~Lewry and J.~Williamson, 
J.~Mat.~Sci. {\bf 29}, 5279-84, (1994). 
A.~J.~Lewry and J.~Williamson,
J.~Mat.~Sci. {\bf 29}, 5524-28, (1994) 
A.~J.~Lewry and J.~Williamson,
J.~Mat.~Sci. {\bf 29}, 6085-90, (1994). 

\bibitem{Bentz91a}\label{Bentz91a}
D.~P.~Bentz and E.~J.~Garboczi, {\em Percolation of phases in a 
three-dimensional cement paste microstructural model}, Cement and 
Concrete Research, vol.~{\bf 21}, 325-44 (1991).

\bibitem{Bentz94a}\label{Bentz94a}
D.~P.~Bentz, P.~V.Coveney, E.~J.~Garboczi, M.~F.~Kleyn and 
P.~E.~Stutzman, {\em Cellular automaton simulations of cement hydration 
and microstructure development}, Modelling Simul.~Mater.~Sci.~Eng.~
{\bf 2}, 783-808 (1994).

\bibitem{Made94}\label{Made94}
B.~Mad\'e, A.~Cl\'ement and B.~Fritz, 
{\em Mod\'elisation thermodynamiqoe et cin\'etique des r\'eactions 
diag\'en\'etiques dans les bassins sedimentaires}, Reveue de L'institut 
francais du petrole {\bf 49} (6), (1994).

\bibitem{thz96}\label{thz96}
F. Tzschichholz, H.J. Herrmann and H. Zanni, 
{\em Reaction-diffusion model for the hydration and setting of cement}, 
Phys. Rev. {\bf E 53}, (3), 2629-37 (1996).

\bibitem{burning}\label{burning}
D. Stauffer, Introduction to Percolation Theory (Taylor and Francis, London, 
1985). 

\bibitem{Scholze85}\label{Scholze85}
H.~Scholze, M.~Hurbaniac and H.~Ruf, 
{\em Comperative considerations on the behaviour of natural gypsum and 
flue gas gypsum}, Zement-Kalk-Gips 10, (1985).

 

\end{thebibliography}
\end{document}